\documentclass[aps,prl,twocolumn,groupedaddress,showpacs]{revtex4}

\usepackage{graphicx}
\usepackage{amsmath}

\begin{document}

\title{Fermi arcs in cuprate superconductors: tracking the pseudogap below $T_{c}$ and above $T^*$}

\author{J.G. Storey$^1$, J.L. Tallon$^{1,2}$, G.V.M. Williams$^{2}$ and J.W. Loram$^3$}

\affiliation{$^1$School of Chemical and Physical Sciences,
Victoria University, P.O. Box 600, Wellington, New Zealand}

\affiliation{$^2$MacDiarmid Institute, Industrial Research Ltd.,
P.O. Box 31310, Lower Hutt, New Zealand}

\affiliation{$^3$Cavendish Laboratory, Cambridge University, Cambridge CB3 0HE, England.}

\date{\today}

\begin{abstract}
Using an energy-momentum dispersion for Bi$_2$Sr$_2$CaCu$_2$O$_8$
obtained from angle-resolved photoelectron spectroscopy we show that
the shrinking Fermi arc model of the pseudogap is inconsistent with
Raman scattering below $T_c$ and specific heat near $T^*$. By
simulating the quasiparticle energy dispersion curves we show that
Fermi arcs are an artifact of a $T$-dependent scattering rate.
\end{abstract}

\pacs{74.25.Gz, 74.25.Jb, 74.62.Dh, 74.72.-h}

\maketitle

The normal-state properties of underdoped
high-$T_{c}$ cuprates are dominated by a gap in the density of
states known as the pseudogap (PG)\cite{PG3,OURWORK1}. Its origin is unknown. Unlike a superconducting gap which \textit{closes} at $T_c$, angle-resolved
photoemission spectroscopy (ARPES) and tunneling in underdoped
cuprates have long suggested that the PG \textit{fills} as $(1-T/T^*)$ and
disappears abruptly at $T^*$\cite{HARRIS,RENNER}. If so then the lost low-energy spectral weight would be restored at
$T^*$ with important thermodynamic consequences, as we shall see.
The gap energy, $E_g$, and $T^*$ decrease with increasing doping,
falling to zero at a critical doping $p=p_{crit}\approx0.19$ holes
per Cu\cite{OURWORK1}. Generally $T^*$ is much greater than $T_c$
but when $p$ exceeds $\approx0.16$ then $T^*$ falls below
$T_c$\cite{NAQIB}.

Recent ARPES studies\cite{FERMIARCS,FERMIARCS2} suggest that for
$T<T^*$ the PG covers only part of the Fermi surface (FS) near the ($\pi$,0) zone boundary. This leaves ungapped arcs on the FS (``Fermi arcs'') which grow with increasing $T$. The arcs extend to an angle $\theta_0$
given by $\theta_0 = \frac{\pi}{4}\left(1-T/T^*\right)$, where
$\theta_0$ is measured from ($\pi$,0). The gap is thus nodal at
$T=0$ and, with increasing $T$, retreats towards ($\pi$,0) where it
closes abruptly at $T^*$. Assuming that the PG continues to exist as
the underlying normal state below $T_c$, the $T$-dependent
restoration of the pristine FS would have important
testable consequences, as follows:

\noindent(i) if the PG retreats abruptly to ($\pi$,0) at $T^*$ the
specific heat coefficient, $\gamma$, will exhibit an anomaly. The
area under the anomaly is exactly equal to the restored entropy. We
calculate this below for several Fermi arc scenarios and show this
is not observed in the experimental data;

\noindent(ii) if the Fermi arcs grow as $T$ rises then the spectral weight taken up by the PG falls. So the superfluid
density would first increase, then fall as $T$ approaches $T_c$,
contrary to the observed monotonic decrease\cite{STOREY};
\begin{figure}
\centering
\includegraphics[width=75mm,clip=true,trim=0 0 0 0]{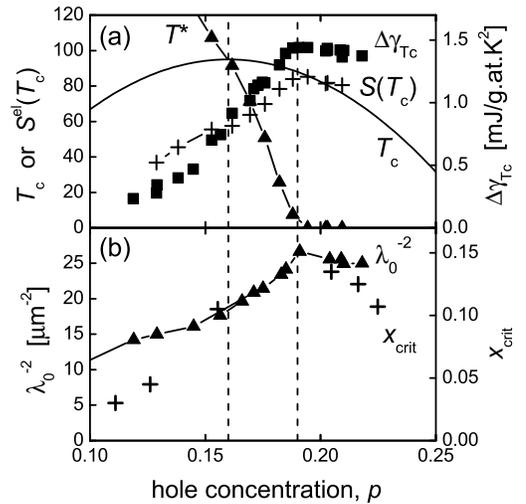}
\caption{Thermodynamic data for Bi-2212 (a) two properties at $T_c$:
the jump in specific heat coefficient $\Delta\gamma$ and the
electronic entropy $S(T_c)$ in mJ/g.at.K; and (b) two properties at
$T=0$: the superfluid density $\lambda_0^{-2}$ and $x_{crit}$ the
critical Zn concentration giving $T_c=0$. Vertical dashed lines
indicate where $T^*=T_c$ and $p_{crit}=0.19$ where the PG closes.}
\label{DELGAMMA}
\end{figure}

\noindent(iii) and if the PG closes at $T^*$ then in the doping range $0.16<p<0.19$, where $T^*\leq{}T_c$, the PG should be open at $T=0$ but closed when $T=T_c$.
Again this is not observed, as we show in Fig.~\ref{DELGAMMA}. Here
$T_c(p)$ and $T^*(p)=E_g/2k_B$ are plotted for
Bi$_2$Sr$_2$CaCu$_2$O$_{8+\delta}$ (Bi-2212). Panel (a) shows two
properties at $T_c$: the jump in
$\Delta\gamma$ and the electronic entropy $S(T_c)$. Panel (b) shows two
ground state $T=0$ properties: the superfluid density
$\lambda_0^{-2}$ and $x_{crit}$, the critical density of Zn required
to suppress $T_c$. The data is from Tallon \textit{et
al.}\cite{TALLONDELGAM}. The two vertical dashed lines show the
region where $T^*\leq T_c$. In (b) both ground-state properties
abruptly reduce when the PG opens at $p_{crit}$. Now if the PG were
to fully close at $T^*$ then the two $T=T_c$ properties displayed in
(a) should not begin to fall until $T^*=T_c$ i.e. at the vertical
line at $p\approx0.16$. They do not. In fact, like the ground-state
properties, they fall abruptly at $p_{crit}=0.19$. This suggests a
relatively $T$-independent loss of spectral weight due to the PG
below $T_c$, impacting roughly equally on ground-state and
$T=T_c$ properties. The Fermi arcs, if they exist, cannot continue
to collapse around the nodes.

Here we explore these points further. We first consider the low-$T$
behavior of the Fermi arcs by calculating the Raman response. Below $T_{c}$ the SC gap opens up on the
Fermi arcs thereby obscuring the details of how the PG further
evolves with $T$, and in particular whether the PG itself becomes
nodal at $T=0$. Raman scattering allows the PG and SC gaps to be separately probed via the
antinodal ($B_{1g}$) and nodal ($B_{2g}$) Raman response\cite{CHEN,OPEL,LETACON}. We calculate these for Bi-2212
using an ARPES derived energy-momentum dispersion. By comparing with Raman data we are able to confirm
that the PG does not evolve significantly below $T_{c}$.

We then turn to the second issue as to how the PG evolves near
$T^*$. Does it really close at $T^*$ as widely believed? We use the same dispersion to calculate the specific heat and
show that the Fermi arc model leads to a large anomaly in
$\gamma(T)$ at $T^*$ that is not observed. We go on
to suggest these issues may be resolved
by incorporating quasiparticle (QP) lifetime broadening.

\begin{figure}
\centering
\includegraphics[width=75mm,clip=true,trim=0 0 0 0]{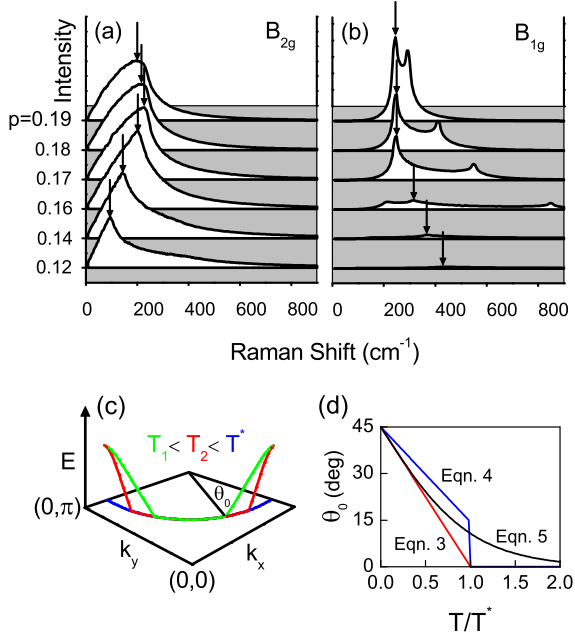}
\caption{(Color online) Calculated Raman response at $T$=0 for a Fermi arc length
fixed below $T_c$: (a) nodal $B_{2g}$ and (b) antinodal $B_{1g}$
symmetry.  Arrows show positions of SC and PG gap features. (c) PG and Fermi arc given by Eqn.~\ref{PGEQ} at temperatures $T_1$ and $T_2$. (d) $\theta_0(T)$ from Eqns. \ref{THETA0EQ}, \ref{THETA0KANIG} and \ref{THETA0EQ2}.}
\label{FIG1}
\end{figure}

We employ the six parameter tight-binding Bi-2212 dispersion,
$\epsilon(\textbf{k})$, reported by Norman \textit{et
al}.\cite{2212EK} and assume a rigid single-band approximation.
Inclusion of band splitting\cite{2212VHS} will not significantly
affect the following results. We take the Fermi level to be 34meV
above the $(\pi ,0)$ van Hove singularity (vHs) near
optimal doping\cite{2212EK} and 96meV above the vHs in an
underdoped sample with $p=0.11$\cite{UD77K}. We interpolate
between for intermediate doping levels. The imaginary part of the
unscreened non-resonant Raman response at $T=0$ is given
by\cite{RAMAN}
\begin{equation}
\chi^{''}_{0}\left(q=0,\omega\right)=\int\frac{dk^2}{\left(2\pi\right)^2}\delta\left(\omega-2E(\textbf{k})\right)\frac{|\Delta(\textbf{k})|^2}{E(\textbf{k})^2}|\gamma(\textbf{k})|^2
\label{RAMANEQ}
\end{equation}
where the integral is over occupied states below $E_F$,
$\Delta(\textbf{k})=\frac{1}{2}\Delta_{0}\left(\cos{k_{x}}-\cos{k_{y}}\right)$
is the $d$-wave SC gap function and
$E(\textbf{k})=\sqrt{\epsilon(\textbf{k})^2+|\Delta(\textbf{k})|^2}$.
In the $B_{1g}$ scattering symmetry
$\gamma(\textbf{k})^{B_{1g}}=\gamma{}B_{1g}\left(\cos{k_{x}}-\cos{k_{y}}\right)$,
giving a dominant response from the antinodal sections of the FS.
For $B_{2g}$,
$\gamma(\textbf{k})^{B_{2g}}=\gamma{}B_{2g}\sin{k_{x}}\sin{k_{y}}$
and the response is mainly nodal. The magnitude of the SC gap,
$\Delta_{0}$, is taken from the weak-coupling result
$2\Delta_{0}=4.28k_{B}T_{c}$ and $T_{c}$ is given by the empirical
relation\cite{OCT1} $T_{c}/T_{c,max}=1-82.6\left(p-0.16\right)^2$.
We adopt a PG of the form
\begin{equation}
E_{g}=\left\{
\begin{array}{ll}
E_{g,max}\cos{\left(\frac{2\pi\theta}{4\theta_{0}}\right)} & (\theta<\theta_{0})\\
\\
E_{g,max}\cos{\left(\frac{2\pi(\theta-\pi/2)}{4\theta_{0}}\right)} & (\theta>\frac{\pi}{2}-\theta_{0})\\
\\
0 & \mbox{otherwise}
\end{array}\right.
\label{PGEQ}
\end{equation}
where $0\leq\theta\leq\pi/2$. We initially assume
\begin{equation}
\theta_{0}=
\begin{array}{ll}
\frac{\pi}{4}\left(1-\frac{T}{T^{*}}\right) & \left(T<T^*\right)
\end{array}
\label{THETA0EQ}
\end{equation}
and $T^{*}=E_{g,max}/k_{B}$. This form of the PG is illustrated in Fig.~\ref{FIG1}(c) and (d). Eqn.~\ref{THETA0EQ} models the linear
$T$-dependence of the Fermi arc length inferred from
ARPES\cite{FERMIARCS2}. At $T$=0, $\theta_{0}=\pi/4$ and the PG is
fully nodal. As $T$ rises, $\theta_{0}$ decreases resulting in a
`filling-in' of the PG and the growth of the Fermi arcs. The PG is a
non-states-conserving gap\cite{ENTROPYDATA2} i.e. unlike the SC gap
there is no pile up of states outside the gap. This is implemented
by removing states with $\epsilon (\textbf{k})<E_{g}$ from the
integration in Eqn.~\ref{RAMANEQ}. The doping dependence of $E_{g}$
is obtained from the reported leading-edge ARPES gap at
100K\cite{OURWORK1}.

\begin{figure}
\centering
\includegraphics[width=80mm,clip=true,trim=0 0 0 0]{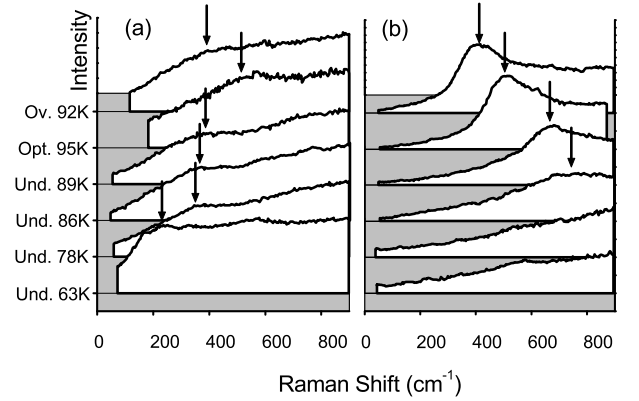}
\caption{Experimental Raman data for Hg-1201 from Le Tacon
\textit{et al}\cite{LETACON}. (a) $B_{2g}$ and  (b) $B_{1g}$
response for $T\ll T_c$.} \label{FIG2}
\end{figure}

Figure~\ref{FIG1}(a) and (b) show the nodal ($B_{2g}$) and antinodal ($B_{1g}$) Raman response
for six dopings spanning the range 0.12 to 0.19. We consider two scenarios:

(i) Firstly, we have assumed that the length of the Fermi arc
becomes fixed at the onset of superconductivity, implemented by
setting $T=T_{c}$ in Eqn.~\ref{THETA0EQ}. Fig.~\ref{FIG1}(a) shows
the nodal ($B_{2g}$) response. Leaving aside the anomalous
electronic Raman continuum above the pairbreaking gap, the
calculations closely resemble the recent results of Le Tacon
\textit{et al}.\cite{LETACON} which are reproduced in
Fig.~\ref{FIG2}. The PG peak maximum in the $B_{1g}$ response shifts
monotonically to higher energies with decreasing doping.
Simultaneously the intensity of this peak rapidly reduces with
underdoping. In contrast, the SC peak maximum in the $B_{2g}$
response is found to shift to lower energies in the underdoped
regime. The magnitude of the $B_{2g}$ peak persists relatively
undiminished down to the lowest doping levels. Also reproduced is
the increased linear slope of the response at very low doping.

\begin{figure}
\centering
\includegraphics[width=75mm,clip=true,trim=0 0 0 0]{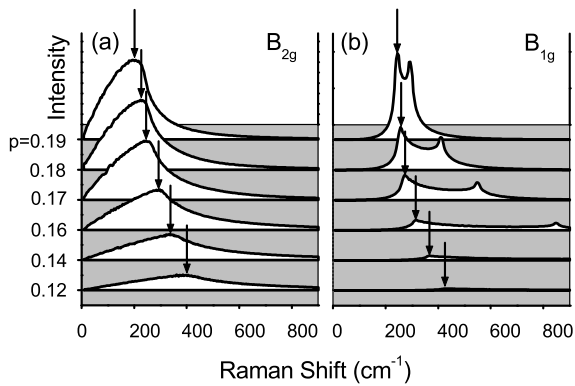}
\caption{Calculated Raman response at $T=0$ with a fully nodal PG
for (a) nodal $B_{2g}$ and (b) antinodal $B_{1g}$ symmetry. The
intensity scale is the same for all plots.} \label{FIG3}
\end{figure}

(ii) Secondly, we show in Fig.~\ref{FIG3} the Raman response in the
alternative case where we have assumed that the Fermi arcs continue
to collapse below $T_c$. This is done setting $T=0$ in
Eqn.~\ref{THETA0EQ} resulting in a fully nodal PG. In this case the
$B_{2g}$ peak shifts monotonically to higher energies with
decreasing doping and the intensity reduces rapidly. This behavior
is not observed experimentally. We cannot say that the Fermi arc
freezes exactly at $T_c$ but our analysis indicates that it remains
finite at $T$=0 and evolves only weakly below $T_c$.

We turn now to the
question as to how the PG evolves above $T_c$ as
$T\rightarrow{}T^*$. It is widely assumed that the PG closes at
$T^*$, as indeed Eqn.~\ref{THETA0EQ} suggests, thus exposing a
pristine FS. In fact, Kanigel \textit{et al}.\cite{FERMIARCS2}
suggest that a discrete jump in $\theta_0$ occurs at $T^*$ so that
the PG retreats to the flat sections of the FS near $(\pi,0)$ from whence it abruptly disappears at $T=T^*$. This model is given
by
\begin{equation}
\theta_{0}=\left\{
\begin{array}{ll}
\frac{\pi}{4}\left(1-0.68\frac{T}{T^{*}}\right) & \left(T<T^*\right)\\
\\
0 & \left(T\geq T^* \right)\\
\end{array}\right.
\label{THETA0KANIG}
\end{equation}
and is illustrated in Fig.~\ref{FIG1}(d).
The closure of the PG, whether according to Eqn.~\ref{THETA0EQ} or
Eqn.~\ref{THETA0KANIG}, will restore the entropy to the bare-band
value. This, quite generally, will result in a $\gamma$ anomaly, the area
of which equals the restored entropy. The lower is $T^*$ the greater
is the anomaly. We evaluate this here.

\begin{figure}
\centering
\includegraphics[width=8cm,clip=true,trim=0 0 0 0]{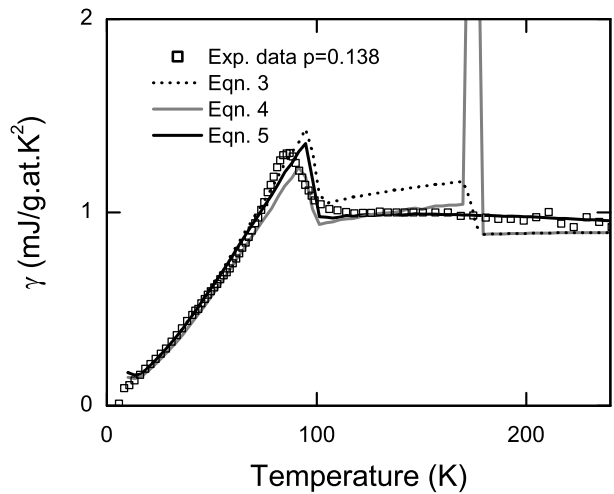}
\caption{Specific heat coefficient calculated for $\theta_0$
given by Eqns. (3), (4) or (5) and compared with the experimental
data for Bi-2212\cite{ENTROPYDATA2} where every 20th data point is shown.}
\label{FIG5}
\end{figure}

Using the method described previously\cite{STOREY}, together with
the above tight-binding dispersion, we have computed $\gamma(T)$ for three cases with
$p\approx0.138$. In Fig.~\ref{FIG5} we compare these with
experimental data for Bi-2212. The three cases are shown in Fig.~\ref{FIG1}(d). They are (i) the
linear behavior described by Eqn.~\ref{THETA0EQ}; (ii) the
sudden jump in $\theta_0$ inferred by Kanigel \textit{et al}.
and described by Eqn.~\ref{THETA0KANIG}; and (iii) the smooth
evolution of the gap given by\cite{STOREY}:
\begin{equation}
\theta_{0}=\frac{\pi}{4}\left(1-\tanh{\left(\frac{T}{T^{*}}\right)}\right)
\label{THETA0EQ2}
\end{equation}
As can be seen, cases (i) and (ii) lead to substantial anomalies in
$\gamma$ which are clearly not found experimentally. The
experimental data, shown by the data points, are from a previous
study on the specific heat of Bi-2212\cite{ENTROPYDATA2} with the
closest corresponding doping state. The smooth evolution of the PG
given by Eqn.~\ref{THETA0EQ2} satisfactorily describes the data and
generally discounts the possibility that the PG closes abruptly at
$T^*$.

We have so far shown that (i) below $T_c$ the Fermi arcs do not
continue to shrink notably; and (ii) above $T_c$ the Fermi arcs
cannot spread out to abruptly form a connected pristine FS at $T^*$.
(iii) Both results are confirmed by the data in Fig.~\ref{DELGAMMA}
in the interesting case where $T^*$ lies below $T_c$. These problems
seriously prejudice the current picture of Fermi arcs but could be
resolved as follows by invoking a $T$-dependent scattering rate.

Firstly, one is led to this view by the observations of Norman
\textit{et al}.\cite{NORMAN}. They modelled the QP peak using a QP
self energy with a scattering rate or inverse lifetime $\Gamma_0$
which, in underdoped samples, grows with $T$ and equals the gap
magnitude precisely at $T^*$ (see Fig. 2(b) of ref. \cite{NORMAN}).
This would lead to a smearing out of the gap at $E_F$ resulting in a
single peak in the dispersion at $E_F$ that looks like a recovered
pristine FS. But of course the gap is still there, and interestingly
is reported to be $T$-independent, just as we concluded above.

\begin{figure}
\centering
\includegraphics[width=8cm]{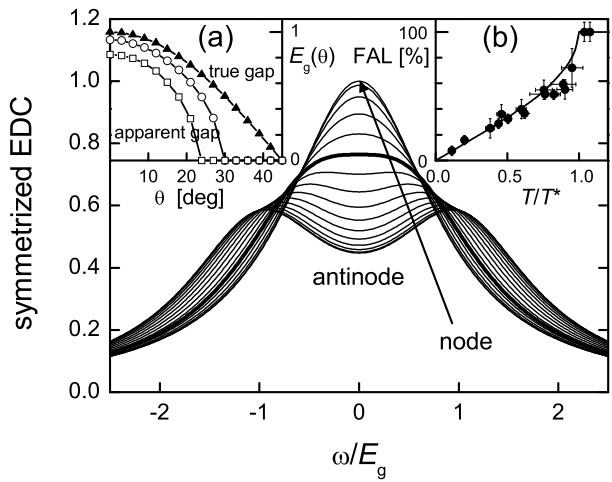}
\caption{\small Simulated symmetrized ARPES quasiparticle EDCs
ranging from the antinode to the node ($\theta$ = 0 to
45$^{\circ}$). The broadening is fixed at 0.45$E_g(0)$. The
crossover from double to single peaks (bold curve) presents an
apparent, though false, closing of the PG and recovery of Fermi
arcs. Inset (a) true normalized gap and apparent gap for two
broadenings (= 0.45$\times$ and 0.6$\times{E_g(0)}$). (b) curve:
apparent Fermi arc length (FAL) assuming broadening $\propto{T/T^*}$; data: from ref. \cite{FERMIARCS2}} \label{FERMIARCS}
\end{figure}

In Fig.~\ref{FERMIARCS} we have simulated the symmetrized QP EDCs as
the sum of two Lorentzians. We use a fixed broadening of
0.45$E_g(0)$, corresponding to a particular temperature below $T^*$.
The $k$-dependent gap value, $E_g(\theta)$ is assumed to be
$d$-wave-like with $E_g(\theta)=E_g(0)\cos(2\theta)$. Fifteen angles
ranging from $\theta$=0 (the antinode) to $\theta$=45$^{\circ}$ (the
node) are shown and the data qualitatively reproduces the reported
experimental data\cite{NORMAN,FERMIARCS2}. These EDCs reveal a
crossover from double to single peaks, with the flat bold curve lying at the
boundary. Within the Fermi arc model this would be interpreted as a
closure of the PG at $\theta\approx 30^{\circ}$ with a pristine
Fermi arc extending from $\theta=30^{\circ}$ to $45^{\circ}$.
But, as shown in inset (a), the \textit{true} gap does not close
until the node at $\theta=45^{\circ}$. The
\textit{apparent} gap, found by reading off the peak positions,
is also plotted in inset (a) and this falls to zero at
$\theta=30^{\circ}$.
We also show the apparent gap for a larger broadening (=0.6$E_g(0)$)
which would correspond to a higher temperature, closer to $T^*$.
Here the apparent gap closes at $23^{\circ}$.

Inset (b) in Fig.~\ref{FERMIARCS} shows the apparent (though
fictitious) Fermi arc length obtained from these simulated EDCs when
the broadening $\Gamma = \Gamma^* (T/T^*)$, where $\Gamma^*$ is the
critical value that ``closes" the gap at the antinode. The arc
length exhibits a rapid change at $T^*$ just like the data of
Kanigel \textit{et al}.\cite{FERMIARCS2} which is also plotted. This
rapid change arises from the flat part of the $d$-wave gap when
$\theta \rightarrow 0$ and does not signal an abrupt recovery of the
FS.

Thus QP lifetime broadening with a $d$-wave gap accounts for all
apparent Fermi arc features, including the abrupt jump in arc length
reflected in Eqn.~\ref{THETA0KANIG}. We expect therefore that the
$T$-variation of the PG is not in Fermi arcs but in the scattering
rate. Both the PG and the SC gap parameters should be replaced by
complex terms of the form
$E_g(\theta)\rightarrow\frac{E_g(\theta)}{1-i\Gamma_1/\epsilon(k)}$ and
$\Delta(\theta)\rightarrow\frac{\Delta(\theta)}{1-i\Gamma_0/\epsilon(k)}$.
This naturally leads to a ``U--shaped" gap\cite{USHAPE2}, and
instead of frozen Fermi arcs below $T_c$ the Raman data would then
insist on a frozen scattering rate.

In summary, we have used an $\epsilon(\textbf{k})$ dispersion and a
model for the normal-state PG, both based on ARPES results, to show
that the shrinking Fermi arc picture is inconsistent with Raman data
below $T_c$ and thermodynamic data near $T^*$. Only by freezing the
length of the Fermi arcs at or near $T_{c}$ do we find that the
calculations mimic the experimental data. This implies that the PG
does not evolve greatly below $T_{c}$. We have further shown that
closure of the PG at $T^*$ and the associated recovery of a pristine
FS would lead to a large specific heat anomaly that is not observed. We thus question the concept of Fermi
arcs and suggest that they are probably an artifact of a
$T$-dependent lifetime scattering rate.

We thank the Marsden Fund for financial support.


\end{document}